This article is a preprint and has not been certified by peer review# Skin marker-based subject-specific spinal alignment modeling: A feasibility study

Stefan SCHMID[a,b,1,*], Lukas CONNOLLY[c,d,e,1], Greta MOSCHINI[c,e,f],
Michael L. MEIER[c,d], Marco SENTELER[c,e,f]

[a]Bern University of Applied Sciences, Department of Health Professions, Division of Physiotherapy,
Spinal Movement Biomechanics Group, Bern, Switzerland
[b]University of Basel, Faculty of Medicine, Basel, Switzerland
[c]University of Zürich, Balgrist University Hospital, Department of Chiropractic Medicine,
Integrative Spinal Research, Zurich, Switzerland
[d]University of Zurich, Zurich, Switzerland
[e]ETH Zurich, Department of Health Science and Technology, Institute for Biomechanics, Zurich, Switzerland
[f]University of Zurich, Balgrist University Hospital, Department of Orthopedics, Zurich, Switzerland
[1]These authors contributed equally

**Correspondence:**
PD Dr. Stefan Schmid, Bern University of Applied Sciences, Department of Health Professions,
Murtenstrasse 10, 3008 Bern, Switzerland, +41 79 936 74 79, stefanschmid79@gmail.com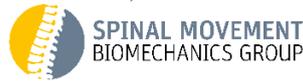

**ABSTRACT**

Musculoskeletal models have the potential to improve diagnosis and optimize clinical treatment by predicting accurate outcomes on an individual basis. However, the subject-specific modeling of spinal alignment is often strongly simplified or is based on radiographic assessments, exposing subjects to unnecessary radiation.

We therefore developed a novel skin marker-based approach for modeling subject-specific spinal alignment and evaluated its feasibility by comparing the predicted with the actual intervertebral joint (IVJ) locations/orientations (ground truth) using lateral-view radiographic images. Moreover, the predictive performance of the subject-specific models was evaluated by comparing the predicted L1/L2 spinal loads during various functional activities with in vivo measured data obtained from the OrthoLoad database.

IVJ locations/orientations were predicted closer to ground truth as opposed to standard model scaling, with average location prediction errors of 0.99±0.68 cm on the frontal and 1.21±0.97 cm on the transverse axis as well as an average orientation prediction error of 4.74°±2.80°. Simulated spinal loads showed similar curve patterns but considerably larger values as compared to in vivo measured data. Differences in spinal loads between generic and subject-specific models become only apparent on an individual subject level.

These results underline the feasibility of the proposed method and associated workflow for inter- and intra-subject investigations using musculoskeletal simulations. When implemented into standard model scaling workflows, it is expected to improve the accuracy of muscle activity and joint loading simulations, which is crucial for investigations of treatment effects or pathology-dependent deviations.

## 1. INTRODUCTION

Musculoskeletal models and computer simulations are increasingly used for investigating biomechanical parameters that are difficult to assess experimentally (Bassani and Galbusera, 2018). The two most commonly used musculoskeletal modeling platforms are OpenSim (Delp et al., 2007) and AnyBody (AnyBody Technology, Arlborg, Denmark). While AnyBody is a commercial software, OpenSim is a free open-source modeling and simulation framework, which includes a growing database of models developed and published by various researchers (Bassani and Galbusera, 2018; Seth et al., 2011). For both platforms, various spine models are available, including models with only articulated lumbar spines (Christophy et al., 2012; de Zee et al., 2007) as well as models with fully articulated thoracolumbar spines and rib cages (Bruno et al., 2015; Ignasiak et al., 2016).

For simulating dynamic activities, these models are usually driven by skin marker-based motion capture data using an inverse kinematics approach (Bassani et al., 2017; Beaucage-Gauvreau et al., 2019; Glover et al., 2021; Ignasiak et al., 2018; Raabe and Chaudhari, 2018). But besides calculating joint angles during dynamic activities, marker data are also used for generic model scaling to account for subject-specific geometries and masses of the individual segments. These scaling procedures are based on a minimization approach, which in almost any case leads to a certain number of residual errors (Glover et al., 2021). It particularly leads to inaccuracies when applied to the spine, which by itself is a multi-joint complex consisting of 24 vertebral bodies, with each having its own dimension and anatomical arrangement. Moreover, the dimensions of the rigid bodies representing the vertebrae are considerably smaller and driven by fewer markers than other body segments such as the extremities. Hence, the definition of spinal alignment based on skin markers represents an underdetermined geometrical problem. On the other hand, the spinal segments (and spinal markers) do not move independently but can be considered a kinematic chain with a large number of degrees of freedom. Consequently, standard personalization strategies in musculoskeletal spine modeling typically compromises subject-specific anatomical differences and neglects skeletal alignments in the spinal region.

To reduce inverse kinematics-based intervertebral joint (IVJ) angle estimation errors without the need for radiographic data, this study aimed at developing and evaluating a novel skin marker-based approach for modeling subject-specific spinal alignment.

## 2. METHODS

### 2.1. Creating full-body models with subject-specific spinal alignment

*2.1.1 Base models*

All models were created using the OpenSim musculoskeletal modeling environment (Delp et al., 2007). As a basis for the development of subject-specific models, we created a set of male and female full-body models, composed of a fully articulated thoracolumbar spine model including a head-neck-complex and upper extremities (Bruno et al., 2015; Bruno et al., 2017) as well as the Gait2354 model (Anderson and Pandy, 1999; Anderson and Pandy, 2001; Delp et al., 1990; Yamaguchi and Zajac, 1989), a standard lower body model available from OpenSim. The models were scaled and combined as previously described (Burkhart et al., 2020; Schmid et al., 2020). To reduce complexity of the model and improve simulation robustness, external and internal intercostal muscle groups were removed, and ribs were considered as rigid bodies attached to the vertebrae. The axial rotation (AR) and lateral bending (LB) coordinates of the thoracic intervertebral joints (IVJs) were locked, reducing them to pin joints with one degree of freedom (flexion and extension (FE)). Additionally, the translational coordinates of the rib to sternum joints were locked, changing them from free to ball joints. To support the muscles (i.e. to account for synergistic passive structures such as the thoracolumbar fascia) and prevent the models from failing in tasks involving large ranges of motion (e.g. lifting up a box from the floor or standing up from a chair) or high impact forces (e.g. running or jumping), coordinate actuators (virtual torque generators) with infinite minimum and maximal control values were added to some joints. The FE, AR and LB coordinates of the L3/L4 to L5/S1 IVJs were thereby actuated with an optimal force of 10 Nm, whereas the remaining lumbar IVJs (T12/L1 to L2/L3) were actuated with an optimal force of 5 Nm and the FE coordinates of the thoracic IVJs with an optimal force of 1 Nm. To further improve simulation robustness, the sternum to clavicle as well as the hip, knee and ankle joints were all actuated with an optimal force of 25 Nm. These optimal forces were defined as large as needed for the model to converge and as small as possible to minimize the effect on the results. To compensate for the dynamic inconsistency between the estimated model accelerations and the measured ground reaction forces, residual actuators were added to all six degrees of freedom of the "pelvis to ground joint", which connects the model to the ground (i.e. to the origin of the coordinate system).

To allow for motion capture data-driven inverse kinematics calculations, we used the drag and drop OpenSim graphical user interface to add virtual markers, in a first step only to the male base model. We thereby chose the full-body marker set described by Schmid et al. (2017) (Figure 1), mainly due to its demonstrated capacity for the reliable assessment of spinal kinematics during functional activities (Niggli et al., 2021; Suter et al., 2020). For all bilateral markers, the locations were mirrored to ensure a symmetric virtual marker set. To add the virtual markers also to the female base model, the marker offsets from the linked body segments were scaled with the same factor as used for scaling the respective body segment from the male to female model.

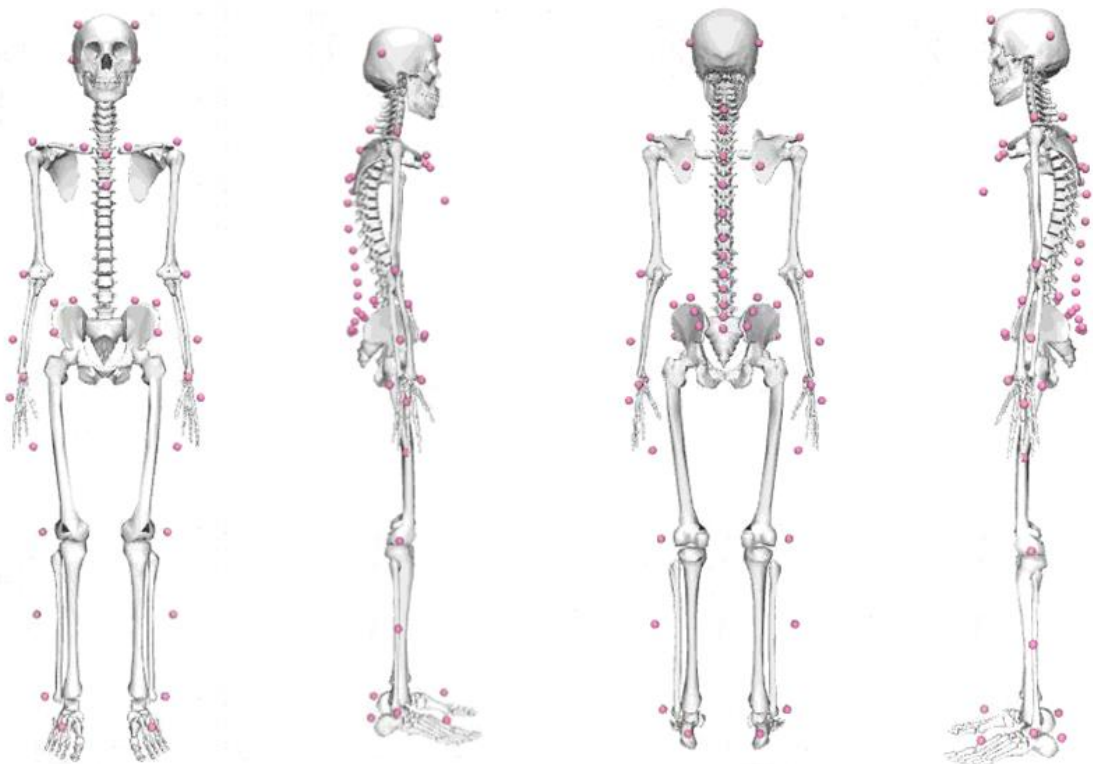

**Figure 1:** The generic male full-body model with the 58 virtual markers according to the configuration described by Schmid et al. (2017).

*2.1.2 Generic model scaling*

To account for individual segmental lengths and masses, base models are first selected according to the gender of the individual under investigation (subject) and scaled generically using the OpenSim scaling tool. This includes scaling of all segments by factors obtained from the distances between marker locations collected during an upright standing trial in the movement laboratory (Table 1). Subsequently, the mass properties of the body segments are scaled while preserving mass distribution and reproducing the total mass (recorded using a standard bathroom scale) of the subject. Muscle properties such as fiber and tendon slack length are then scaled proportionally to the total muscle resting length. Additionally, virtual marker locations are adapted to match the experimental marker locations using the OpenSim scaling tool option 'Adjust Model Markers'.

**Table 1:** Reference marker pairs for generic model scaling. Marker names as well as marker placement locations correspond to the configuration described by Schmid et al. (2017).

| Model region | Reference Marker Pairs | Scaled Bodies (X, Y, Z) |
|---|---|---|
| Torso | C7 → SACR (S2) | Pelvis, Sacrum, Abdomen, all Intervertebral Bodies, all Ribs, Sternum, Clavicles, Scapularies, Head / Neck |
| Upper extremities | (R/L)SHO → (R/L)FIN | Humerus, Ulna, Radius, Hand (all bilateral) |
| Lower extremities | (R/L)ASI → (R/L)ANK<br>(R/L)THI → (R/L)ANK | Femur, Tibia, Talus, Calcaneus, Toes (all bilateral) |

*2.1.3 Implementation of skin marker-derived spinal alignment*

Following generic model scaling, spinal alignment is adjusted based on the external back profile of the subject using a custom MATLAB routine (R2020b, MathWorks, Inc., Natick, MA, USA). First, a 4th order polynomial is fitted through the sagittal plane coordinates of the markers placed over the spinous processes of C7, T3, T5, T7, T9, T11, L1-L5 and the sacrum (height of S2) (Figure 2A). The sagittal plane is thereby defined by a plane fit into the locations of the 12 spinal markers (C7 to sacrum) as well as the markers placed on the sternum and bilaterally on claviculae. Through fitting the polynomial, potential discontinuities arising for example from marker (mis)placement are smoothed out by setting the horizontal (anterior-posterior) position of each marker as a result of the polynomial function of vertical (superior-inferior) position. Input points for the internal spinal alignment polynomial are then acquired by subtracting the scaled marker to joint center distance (in the base model) from the smoothened marker position. The underlying assumption thereby is a vertebral orientation along the normal vector of the external back profile polynomial at the height of the corresponding experimental marker (Figure 2B). Exceptions are the predicted IVJ locations for the segments C7/T1 and T1/T2, since the "head and neck segment" (to which the C7 marker is linked to) is always oriented horizontally, and therefore, the T1/T2 IVJ location is calculated based on its offset from the C7 marker.

The sagittal plane projection of the geometric center between the two posterior head markers is added as a supporting point to account for an appropriate cervical lordosis. These points are then used to fit a second 4th order polynomial for predicting the internal spinal alignment (Figure 2C). The estimated joint center locations in the sagittal plane are derived from the polynomial by setting the vertical position of each IVJ in the scaled base model as function input (Figure 2D). The orientation of each vertebrae is derived from the angle of the normal vector at the joint center with respect to the frontal axis (Figure 2E). Subject-specific pelvis orientation was not accounted for and therefore, sacrum and pelvis inclination of the subject-specific models remained the same as in the base models.

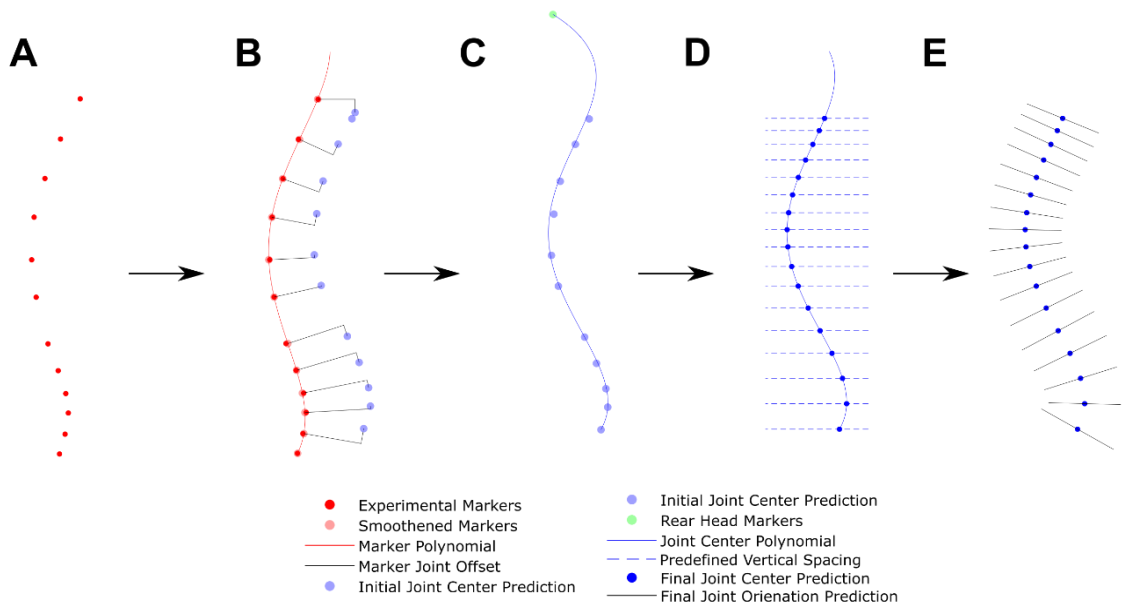

**Figure 2:** Subject-specific spinal alignment adjustment. A) Experimental skin marker locations attached over the respective spinous processes. B) Approximation of intervertebral joint center locations (blue dots) by subtracting the marker-joint offsets from the external marker positions (red dots). Marker-joint offsets were obtained from virtual markers and joint centers in the generic model, after scaling the model to match the subject dimension. Underlying assumption was a vertebral orientation according to the normal vector of the external back profile (4th order polynomial in red). C) Internal spine shape obtained from 4th order polynomial (in blue) by using predicted joint centers and a supporting posterior head marker center point. D) Final predicted joint centers with predefined relative vertical spacing (according to the generic model) between intervertebral joints. E) Final calculated joint orientation determined by the normal vector to the polynomial representing subject-specific spinal alignment.

The estimated IVJ locations and orientations are then converted from global to local three-dimensional coordinate systems. Since the joint coordinate systems of the template model coincide with the centroids of the corresponding vertebral bodies, location and orientation of each joint center are calculated with respect to the former joint in a global frame and converted to a local coordinate system by fixing each joint center in space and rotating the reference frame by its orientation angle. Additional model modifications such as rib and sternum relocation/reorientation are made to maintain the global distance and orientation from the base model to the modified spine. Center of mass locations for each spinal segment are readjusted to ensure the same relative position to the vertebral body. And finally, muscle fiber and tendon slack length are adjusted to match the proportions of the base model.
The OpenSim base models as well as the MATLAB code for subject-specific model scaling and spinal alignment adjustment are freely available via the project-hosting platform SimTK (https://simtk.org/projects/pers_fbm_spine).

## 2.2 Feasibility evaluation

Feasibility was evaluated by two separate studies, which both were approved by the local ethics committee and for which all participants provided written informed consent.

*2.2.1 Spinal alignment adjustment*

To evaluate the subject-specific adjustment of spinal alignment, we conducted a radiographic analysis with three healthy individuals (2 females, 1 male; age: 27±7.5 years; mass: 64.3±13.7 kg; height: 173.7±5.7 cm). We inserted radio-opaque metallic spheres (fiducial) with a diameter of 3.5mm into standard retro-reflective skin markers and placed them as described above over the spinous processes of C7, T3, T5, T7, T9, T11, L1-L5 and the sacrum. Each of the three individuals then underwent a full-body upright standing biplanar radiographic examination using a low dose clinical x-ray scanner (EOS; EOS Imaging, Paris, France). We then manually identified the four corner points of each vertebral body as well as the center point of each skin marker on the lateral view images using a custom MATLAB-based graphical user interface. The ground truth joint center position (i.e. the geometric center of the intervertebral disc in the sagittal plane) for each spinal segment was established by averaging the lower two corner points of the cranial vertebra and the upper two corner points of the caudal vertebrae, whereas the skin marker-derived joint center positions were predicted as described in section 2.1.3. To estimate the accuracy of the spinal alignment modeling approach, we calculated the differences between ground truth and the predicted joint centers for all three subjects as well as the mean errors between subjects in terms of position (frontal and transverse axis), vertebral body orientation and the two commonly used spinal shape parameters lumbar lordosis (LL) and thoracic kyphosis (TK).

*2.2.2 Simulation of spinal loading during various functional activities*

To evaluate the suitability of our subject-specific models for simulating spinal loading during various functional activities, we predicted the compressive forces for the L1/L2 segment during various functional activities and qualitatively compared them to the L2 instrumented vertebral body replacement (VBR) measurements available from the OrthoLoad database (https://orthoload.com).

We therefore recruited 17 healthy individuals (13 males, 4 females; age: 28±4.95 years; mass: 76.6±10.4 kg; height: 179.3±7.84 cm) and invited them for a single visit to our movement laboratory. After the acquisition of relevant anthropometric data (body mass, height), they were equipped with 58 retro-reflective skin markers in a configuration corresponding to the one implemented in the model (see section 2.1.1 and Figure 1) and asked to stand upright for 10 seconds as well as to perform the functional activities walking, running (not evaluated in this study), lifting a 5 kg-box (box lifting), standing up from a chair (chair rising) and climbing stairs. Apart from standing upright, all tasks were performed five times. Motion data were recorded at a sampling rate of 200 Hz using a 27-camera optical motion capture system (Vicon, Oxford, UK). Recorded data were processed using the software Nexus (version 2.8.1, Vicon, Oxford, UK) and transformed into the OpenSim reference system using a custom MATLAB routine. Ground reaction forces were recorded at a sampling rate of 1 kHz using six level ground embedded force plates (AMTI, BP400600) and two mobile force plates (Kistler, 9260AA6) mounted on the first and second step of a four-step staircase.

Following data acquisition, subject-specific models were created based on the mean marker locations obtained during the first 5 seconds of the upright standing trial (see sections 2.1.2 and 2.1.3). For the box lifting task, a separate set of models was created by adding 2.5 kg to each hand and updating the inertial properties to account for the measures of the respective side of the box. No other model properties were changed for the box lifting task. For the walking and stair climbing tasks, start and stop events were set manually using the software

Nexus, whereas for the chair rising and box lifting tasks, a custom MATLAB-based event detection algorithm was used.

Simulations of spinal loading were carried out in three steps: 1) Inverse kinematics calculations to obtain joint angles at each timepoint by solving a weighted least squares error problem. 2) Static optimization to estimate activity and force of the individual muscle fibers by minimizing the sum of squared muscle activations (Herzog, 1987). 3) Joint reaction force estimations by solving the static equations for each timepoint and joint using the estimated muscle forces from the static optimization, experimental ground reaction forces and segmental masses.

## 3. RESULTS

Our approach was able to estimate IVJ center locations closer to ground truth as compared to standard model scaling (Figure 3). The absolute average mean IVJ center location prediction errors were 0.99±0.68 cm on the frontal and 1.21±0.97 cm on the transverse axis, with maximum values reaching 2.47 cm (T8/T9 joint of Subject 2) and 3.40 cm (T1/T2 joint of Subject 3), respectively (Figure 4). The absolute average mean vertebral body orientation prediction error was 4.74°±2.80°, with a maximum value of 11.97° for the L5/S1 joint of Subject 2. Prediction errors for lumbar lordosis and thoracic kyphosis angles were 6.26°±9.51° and 7.33°±3.39°, respectively.

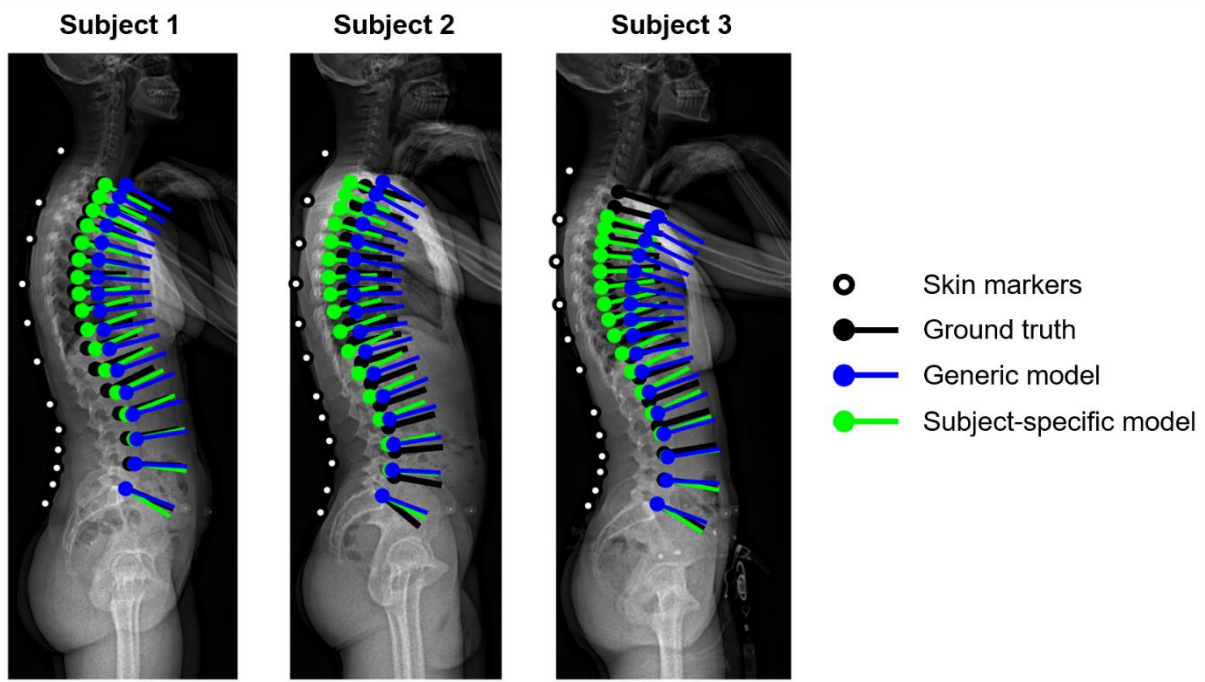

**Figure 3:** Radiographic lateral-view images of the three subjects used for the evaluation of the subject-specific spinal alignment adjustments. The black dots and lines represent the intervertebral joint center locations and orientations derived from the radiographs (ground truth), while the blue and green ones represent the joint center locations and orientations of the generic model and the subject-specific adjustments, respectively.

Mean compressive forces at the L1/L2 level showed no obvious differences between the generically scaled models and the models with the subject-specific spinal alignment (Figure 5). Compared to the in vivo measured OrthoLoad data, however, the predicted compressive forces showed similar shapes but considerably larger values over all the simulated activities.

For walking and stair climbing, model predictions were on average 40-150% larger, with normalized loads reaching values of 10 and 15 N/kg, respectively. Predicted loads for the chair rising task were on average up to 95% larger during the most relevant part of the rising phase, with average peak values reaching approximately 20 N/kg. The box lifting simulations predicted loads that were on average up to 70% larger during most of the lifting-up phase, with average peak values reaching approximately 32 N/kg.

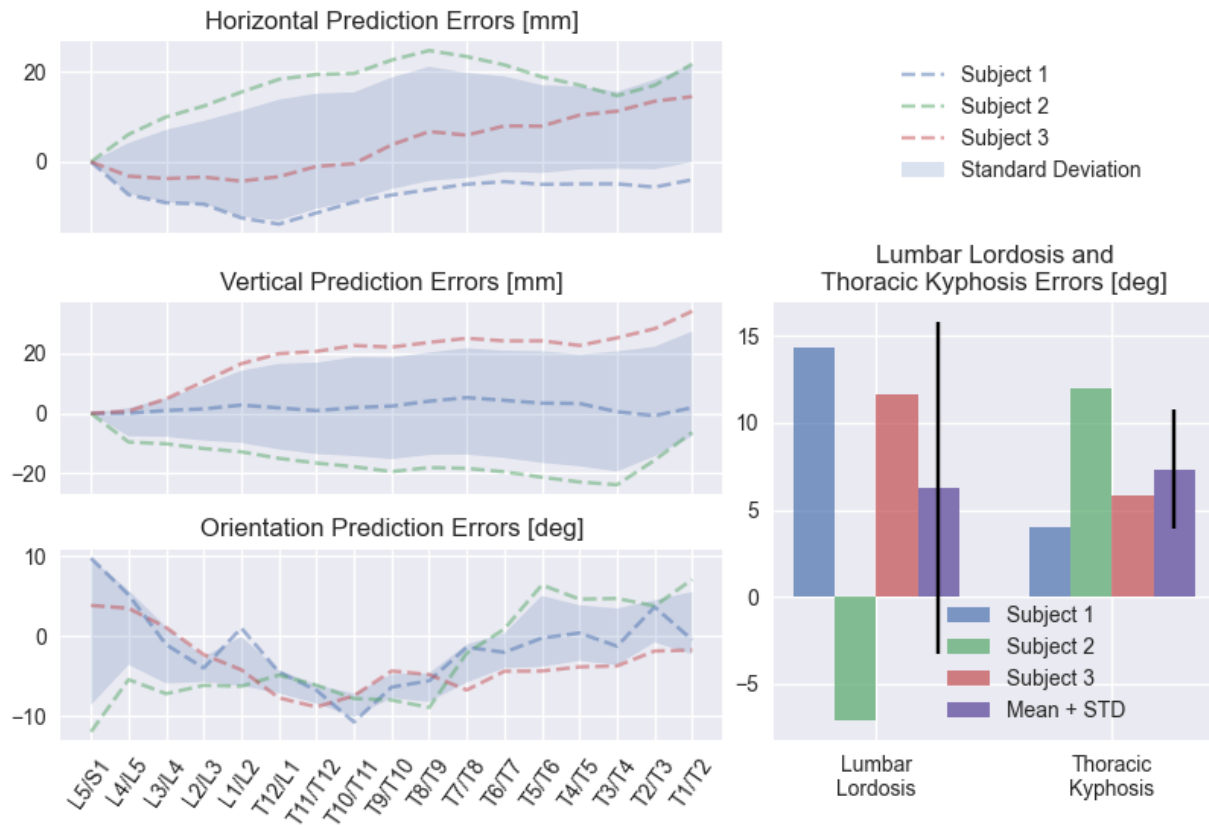

**Figure 4:** Prediction errors for the intervertebral joint center locations and orientations (left side) as well as for the lumbar lordosis and thoracic kyphosis angles (right side) of the three evaluated subjects.

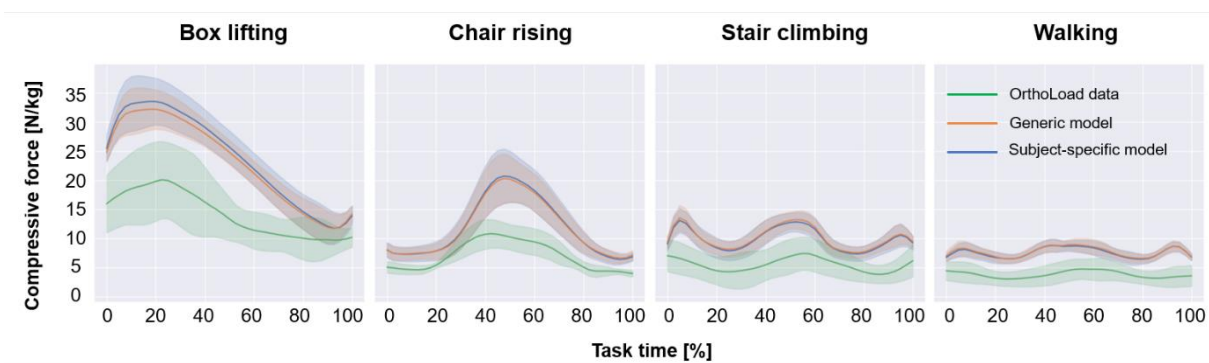

**Figure 5:** Compressive joint forces at the L1/L2 level during box lifting, chair rising, stair climbing and walking, derived from the OrthoLoad database (green lines) as well as resulting from simulations with the generic (orange lines) and subject-specific models (blue lines). The shaded areas represent the standard deviations.

## 4. DISCUSSION

This study introduces a novel skin marker-based approach for modeling subject-specific spinal alignment, which enables subject-specific simulations of trunk muscle activity and spinal loading during various functional activities. Radiographic evaluations during upright standing revealed that the proposed approach was able to predict IVJ center locations/orientations closer to ground truth as opposed to standard model scaling, with average joint center location prediction errors of 0.99±0.68 cm on the frontal and 1.21±0.97 cm on the transverse axis as well as an average vertebral body orientation prediction error of 4.74°±2.80°. Simulations of L1/L2 intervertebral compressive forces during functional activities showed similar curve patterns but considerably larger values as compared to in vivo measured data available from the OrthoLoad database.

Even though the proposed approach resulted in a more accurate prediction of internal spinal shape, it still heavily relies on the distances between markers and joint centers of the base models, which were simply set by drag and drop in the OpenSim graphical user interface and without considering the individualism among subjects with respect to these distances. In a next step, it might therefore be meaningful to implement linear regression models for a better approximation of the distances between markers and joint centers by variables such as gender, height, mass and age, and thus increase the subject-specificity of the model and potentially further improve IVJ center location and vertebral orientation predictions. The relatively large variation in predicted IVJ center locations is likely a direct consequence of this limitation, despite all subjects being in a rather normal range of BMI and posture. A more sophisticated inclusion of marker-to-joint distance estimations already in the base model would hence be advised before generalizing the method. This particularly applies to modeled deformities or other conditions such as obesity.

Furthermore, despite the subject-specific spinal alignment adjustments, the models created with this method remain generic to some extent. To prevent large errors caused by inaccurate marker placements, the method does not adjust the vertical spacing of the vertebrae, which resulted in some significant prediction errors. The use of two markers (C7 and S2) for scaling the whole torso including the length of the spine might have led to scaling errors, as demonstrated by "Subject 3" in the radiographic evaluation study. Using more markers to calculate the scaling factor might improve spinal height approximation, but the prediction of subject-specific vertebral morphometric features such as disc height, vertebral body height and transverse width will remain generic. Joint orientations are also generic by being always perpendicular to the spline at the corresponding joint location, as it is impossible to reliably derive orientations based on fewer skin markers than joints. From a computational point of view such a simplification has even advantages, as, similar to smoothening marker positions by the polynomial fitting, they ensure a certain amount of continuity in the arrangement of the spinal column, which increases the computational robustness of created models.

The fact that the radiographic evaluation of our approach was only based on three healthy and relatively young subjects was considered a major limitation. Moreover, the spinal posture during upright standing in the EOS system might have been slightly different from the one in the movement laboratory. Although this issue was previously encountered (Schmid et al., 2015), we were not able to conduct the two measurements simultaneously. Future evaluations are therefore encouraged to include larger sample sizes with a broader range of subjects, and to conduct simultaneous measurements.

The almost identical L1/L2 mean compressive forces resulting from the generically scaled models and the models with subject-specific spinal alignments gives the impression that the spinal alignment adjustments did not have any effect on intervertebral loading. However, when looking at the curves within the individual subjects, it becomes apparent that only a few subjects showed almost identical forces, while the others showed either slightly lower or higher forces (Figure 6).

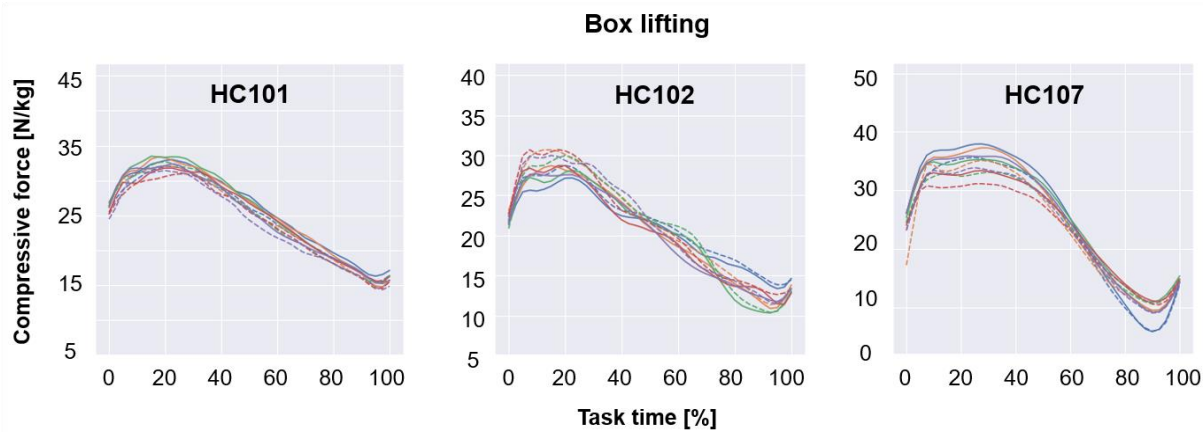

**Figure 6:** Exemplary compressive joint forces at the L1/L2 level during box lifting resulting from simulations with the generic (dotted lines) and subject-specific models (solid lines). Forces were either almost identical (subject "HC101"), slightly lower (subject "HC102"), or slightly higher (subject "HC107").

Our models were able to estimate joint loading for the activities walking, running, chair rising, box lifting and stair climbing, without any need for task-specific modifications such as described in Beaucage-Gauvreau et al. (2019). However, the robustness of our models comes at the cost of adding numerous coordinate actuators (artificial torque generators) to the spinal joints, which provide support when the optimizer is unable to find a solution based on the modeled muscles. The torques generated by these actuators are not negligible, but in comparison with muscle-generated torques, their implementation seems adequate. During model development we ensured that the actuation limits were kept as low as possible, just high enough to ensure model convergence while avoiding excessive torques. Particularly with respect to missing an explicit soft-tissue representation in the model, the torque actuators can be considered a way to account for highly variable passive stiffness properties within the musculoskeletal system. Still, the lack of explicitly modeled soft tissue and negligence of other factors such as intra-abdominal pressure might partially explain the differences in joint loading observed between our model predictions and the in vivo measured data. Another possible explanation is the design of the instrumented implants, particularly the external rods, suggesting that the sensors did not account for the entire load transferred through the vertebral body. On the other hand, our models did not include facet joints, which means that the entire load was transferred through the intervertebral disc. It can therefore be assumed that the "real" intersegmental loads probably lie somewhere in between the in vivo measurements and our model predictions.

In conclusion, we created OpenSim male and female musculoskeletal full-body models, which enable simulations of various functional activities without the need for task-specific modifications. Moreover, we introduced a novel skin marker-based approach for subject-specific spinal alignment adjustments. Both models and code are freely available via the project-hosting platform SimTK. When implemented into standard model scaling workflows, it is expected to improve the accuracy of muscle activity and joint loading simulations, which is crucial for investigations of treatment effects or pathology-dependent deviations.

## 5. CONFLICT OF INTEREST STATEMENT

The author declares no conflict of interest.

## 6. ACKNOWLEDGMENTS

This research was supported by the Swiss National Science Foundation (SNSF, Bern, Switzerland). Movement analysis was performed with support of the Swiss Center for Clinical Movement Analysis (SCMA). The authors thank Marina Hitz, Linard Filli, and Marc Bolliger from the SCMA for their support.